\documentclass{aa}

\usepackage{graphicx}
 \graphicspath{{./Pics/}}
\usepackage{txfonts}
\usepackage{natbib}

\newcommand{\beq}{\begin{equation}}
\newcommand{\eeq}{\end{equation}}
\newcommand{\beqa}{\begin{eqnarray}}
\newcommand{\eeqa}{\end{eqnarray}}

\newcommand{\un}[1]{{\,{\rm #1}}}
\newcommand{\dw}[1]{_{\scriptstyle \mathrm{#1}}}

\defcitealias{2007A&A...469...11H}{Paper II}

\begin{document}

\title{Flow instabilities of magnetic flux tubes}
\subtitle{III. Toroidal flux tubes}
\titlerunning{Flow instabilities of toroidal flux tubes}
\author{V. Holzwarth} 
\offprints{V. Holzwarth} 
\institute{Max-Planck-Institut f\"ur Sonnensystemforschung,
Max-Planck-Strasse 2, 37191 Katlenburg-Lindau, Germany \\
\email{holzwarth@mps.mpg.de}}

\date{Received ; accepted}

\abstract
{The stability properties of toroidal magnetic flux tubes are relevant
for the storage and emergence of magnetic fields in the convective
envelope of cool stars.
In addition to buoyancy- and magnetic tension-driven instabilities, 
flux tubes are also susceptible to an instability induced by the
hydrodynamic drag force.}
{Following our investigation of the basic instability mechanism in the 
case of straight flux tubes, we now investigate the stability
properties of magnetic flux rings. 
The focus lies on the influence of the specific shape and equilibrium 
condition on the thresholds of the friction-induced instability and on
their relevance for emerging magnetic flux in solar-like stars.}
{We substitute the hydrodynamic drag force with Stokes law of friction to 
investigate the linear stability properties of toroidal flux tubes in
mechanical equilibrium.
Analytical instability criteria are derived for axial symmetric
perturbations and for flux rings in the equatorial plane by analysing
the sequence of principal minors of the coefficient matrices of
dispersion polynomials.
The general case of non-equatorial flux rings is investigated numerically 
by considering flux tubes in the solar overshoot region.}
{The friction-induced instability occurs when an eigenmode reverses its
direction of propagation due to advection, typically from the
retrograde to the prograde direction.
This reversal requires a certain relative velocity difference between
plasma inside the flux tube and the environment.
Since for flux tubes in mechanical equilibrium the relative velocity
difference is determined by the equilibrium condition, the instability
criterion depends on the location and field strength of the flux ring.
The friction-induced instability sets in at lower field strengths than
buoyancy- and tension-driven instabilities.
Its threshold is independent of the strength of friction, but the
growth rates depend on the strength of the frictional coupling between
flux tube and environment.}
{The friction-induced instability lowers the critical magnetic field
strength beyond which flux tubes are subject to growing perturbations.
Since its threshold does not depend explicitely on the friction
parameter, this mechanism also applies in case of the quadratic
velocity dependence of the hydrodynamic drag force.
Whereas buoyancy- and tension-driven instabilities depend on the
magnetic field strength alone, the dependence of hydrodynamic drag on
the tube diameter gives rise to an additional dependence of growth
times on the magnetic flux.}

\keywords{Magnetic fields -- Magnetohydrodynamics (MHD) --
Instabilities -- Sun: magnetic fields -- Stars: magnetic fields}

\maketitle

\section{Introduction}
The magnetic field observed on the surfaces of cool, solar-like stars 
is assumed to originate in the bottom of the stellar convection
zone.
It is amplified by the shear flow in the tachocline and stored in the
stably stratified overshoot region at the interface to the radiative
core.
The toroidal magnetic fields are subject to perturbations through
overshooting gas plumes penetrating from the convection zone above.
Beyond a critical field strength, unstable perturbations lead to
growing magnetic flux loops, which rise through the convection zone and
eventually emerge at the stellar surface
\citep[e.g.][]{1982A&A...106...58S, 1996A&A...314..503S}.
In the solar overshoot region, critical field strengths are of the
order of $10^5\un{G}$ \citep[e.g.][]{1993GAFD...72..209,
1995GAFD...81..233}.
Simulations of rising flux tubes with such initial field strengths
yield emergence properties which are consistent with observed
properties of sunspots such as their eruption latitudes, tilt angles,
proper motions, and asymmetries between the leading and the following
component of bipolar spot groups \citep{1993A&A...272..621D,
1994ApJ...436..907F, 1994SoPh..153..449M, 1995ApJ...441..886C}.

The stability properties of magnetic flux tubes are relevant for the
amplification, storage, and emergence of magnetic fields.
An efficient generation of magnetic flux requires the field to remain
within the convection zone for time periods which are comparable with
or longer than the amplification timescale of the (here, unspecified)
dynamo mechanism.
Instability mechanisms limit the storage time and lead to their leakage
of magnetic fields from the overshoot region with field-dependent
growth times.
The main driving mechanism of unstable, non-axisymmetric perturbations
is magnetic buoyancy, which gives rise to undulatory, Parker-type
instabilities \citep{1955ApJ...121..491P}.
If the flux tube dynamics is dominated by the magnetic tension force,
an axial-symmetric poleward slip instability can occur
\citep{1982A&A...106...58S}.

The hydrodynamic drag force reduces the relative motion of a flux tube
perpendicular to its environment, similar to the case of a solid
cylinder immersed in an external flow.
The intrinsically dissipative interaction with the environment can
cause under certain conditions the onset of flux tube instabilities
\citep[e.g.][]{1988JETP...67.1594R, 1997SoPh..176..285J}.
A friction-induced instability occurs if the relative flow velocity
along the magnetic flux tube is higher than a critical value.
In paper II of this series \citep{2007A&A...469...11H}, we have
investigated friction-induced instabilities of straight, horizontal
flux tubes with parallel flows to elucidate the basic instability
mechanism.
In that case, the velocity difference between the internal and external
plasma was a free parameter.
Here we consider magnetic flux rings in a stratified environment to
investigate the influence of the toroidal geometry and of the
equilibrium condition on the stability properties.
Both aspects affect the velocity difference between the internal and
external plasma.

In Sect.~\ref{model} we describe the linearisation of the equations of
motion in the framework of the thin flux tube-model, including a
frictional interaction with the environment.
In the linear stability analysis in Sect.~\ref{linsa}, analytical
instability criteria are derived for some special cases.
The general case is analysed numerically on the basis of flux rings
located in the solar overshoot region.
Section \ref{disc} contains the discussion of the results and
Sect.~\ref{conc} the conclusions.

\section{Model description}
\label{model}

\subsection{Thin flux tube approximation}
The stability analysis is carried out in the framework of the thin flux
tube approximation \citep{1981A&A...102..129S}.
The radius of the flux tube is small compared with the radius of
curvature, the wavelength of a perturbation, and the pressure scale
height.
Cross-sectional variations are taken to be negligible.
All quantities are given by their values on the tube axis and are 
functions of time, $t$, and arc length, $s$, only.
The flux tube is assumed to always retain a circular cross section.

The dynamics of the magnetic flux tube is determined through the
equation of motion
\beq
\rho
\frac{d}{dt} \vec{u}
=
-
\nabla \left(p+\frac{\vec{B}^2}{8\pi}\right)
+
\rho \vec{g}
+
\vec{n}
\frac{\vec{B}^2}{4\pi}\,\kappa 
+ 
\vec{t}
\frac{\partial}{\partial s} \frac{\vec{B}^2}{8\pi}
+
2 \rho \left(\vec{u}\times\vec{\Omega}\right)
+
\vec{f}\dw{D}
\ ,
\label{inteom}
\eeq
where $\vec{u}$ is the plasma velocity, $\vec{B}$ is the magnetic
field, $\rho$ is the density, $p$ is the gas pressure, $\kappa$ is the
curvature of the tube, and $\vec{\Omega}$ is the constant rotation
vector of the star.
For most but the fastest rotating stars the contribution of the
centrifugal acceleration to the effective gravitation is small.
We therefore take $\vec{g}\simeq - g \vec{e}_r$, with $\vec{e}_r$ being
the radial unit vector in a spherical coordinate system and $g\,(=
\textrm{const})$ being the local gravitation.
The adopted coordinate system is the tripod co-moving with the magnetic
flux tube (Frenet basis), consisting of the tangential vector
$\vec{t}$, the normal vector $\vec{n}$, and the binormal vector
$\vec{b}$.
The tangential vector defines the axis of the flux tube, implying
$\vec{B}= B \vec{t}$.

\subsection{Drag force}
In the limit of infinite conductivity, plasma cannot cross the magnetic
surface of a flux tube.
Consequently, a motion of the flux tube perpendicular to its tube axis
implies a flow of the external plasma around the tube.
Analogous to a cylinder immersed in a non-ideal streaming fluid, this
flow gives rise to a hydrodynamic drag force
\beq
\vec{f}\dw{D}
=
-
\rho\dw{e}
\frac{C\dw{D}}{\pi a}
\left| \vec{u}\dw{rel,\perp} \right|
\vec{u}\dw{rel,\perp}
\ ,
\label{defdrag}
\eeq
where $\rho\dw{e}$ is the density of the external plasma, $a$ is the
tube radius, $C\dw{D}$ is the (empirical) drag coefficient, and
$\vec{u}\dw{rel,\perp}$ is the component perpendicular to the tube axis
of the velocity difference between the internal and the external
plasma.
In the reference frame co-rotating with the star, the external plasma
is at rest and the relative perpendicular flow velocity thus given by
\beq
\vec{u}\dw{rel,\perp}
= 
\vec{u} - \left( \vec{u} \cdot \vec{t} \right) \vec{t}
\ .
\label{defurel}
\eeq
In stationary equilibrium, the plasma flow inside the flux tube is
parallel to the tube axis.
Owing to its quadratic dependence on the velocity, the hydrodynamic
drag force drops out of the linearised equation of motion
\citep[e.g.][]{1993GAFD...72..209}.
To investigate the principal influence of friction on the stability
properties of magnetic flux tubes, we consider the case of a linear
velocity dependence according to Stokes's law:
\beq
\vec{f}\dw{D}\rightarrow 
\vec{f}\dw{St}
= 
- \rho\dw{e} \alpha \vec{u}\dw{rel,\perp}
\label{frictrans}
\ .
\eeq
The friction parameter, $\alpha> 0$, is taken to be constant.

\subsubsection{External stratification}
Disregarding any differential rotation or meridional circulation, the
stellar stratification in the co-rotating reference frame is determined
through
\beq
\nabla p\dw{e}
=
\rho\dw{e}\,\vec{g}
\ ,
\label{exteom}
\eeq
where $p\dw{e}$ and $\rho\dw{e}$ are the external gas pressure and
density, respectively.
The flux tube is in lateral pressure balance with its field-free
environment:
\beq
p\dw{e}= p + \frac{B^2}{8\pi}
\ .
\eeq
Subtracting Eq.\ (\ref{exteom}) from Eq.\ (\ref{inteom}) yields 
\beqa
\frac{d}{d t} \vec{u}
& = &
\frac{\Delta\rho}{\rho}
\vec{g}\dw{eff}
+
\vec{n}
\kappa
\frac{B^2}{4\pi\rho}
+ 
\vec{t}
\frac{1}{\rho}
\frac{\partial}{\partial s}
\frac{B^2}{8\pi}
+
2
\left(
  \vec{u}
  \times
  \vec{\Omega}
\right)
- 
\frac{\rho\dw{e}}{\rho}
\alpha 
\vec{u}\dw{\perp}
\label{starteq}
\ ,
\eeqa
where $\Delta\rho= \rho - \rho\dw{e}$ is the density contrast.
Equation (\ref{starteq}) is the starting point for the linear stability
analysis.

\subsection{Linearised equation of motion}

\subsubsection{Equilibrium condition}
We consider magnetic flux rings in mechanical equilibrium located
parallel to the equatorial plane.
A detailed description of this equilibrium and how it is obtained is
given by \citet{1992A&A...264..686M} and \citet{1995ApJ...441..886C}.

In the following, the index `0' indicates equilibrium values.
Since the equilibrium flux ring is axially symmetric, the tangential
component of Eq.\ (\ref{starteq}) vanishes.
From the binormal component follows that the density contrast is nil,
because buoyancy provides the only force component parallel to the
rotation axis.
With $\Delta \rho\dw{0}= 0$ the normal component of Eq.\
(\ref{starteq}) yields
\beq
u_0^2
+ 
2 u_0 v\dw{e,0}
-
c\dw{A,0}^2
=
0
\ ,
\label{equi}
\eeq
where $c\dw{A,0}= B_0/\sqrt{4\pi\rho_0}$ is the Alfv\'en velocity and
$v\dw{e,0}= \Omega/\kappa\dw{0}$ is the rotation velocity of the tube's
environment.
Equation (\ref{equi}) determines the velocity,
\beq
u_0
=
v\dw{e,0}
\left(
 \sqrt{ 
  1 
  + 
  \left( \frac{c\dw{A,0}}{v\dw{e,0}} \right)^2
 }
 -
 1
\right)
>
v\dw{e,0}
\ ,
\label{defuequi}
\eeq
of the plasma inside the flux ring which is required to balance the
magnetic tension force through Coriolis and inertial forces.
This internal plasma flow is in prograde direction.
We shall refer to the dimensionless quantity
\beq
\eta
=
\frac{c\dw{A,0}}{v\dw{e,0}}
=
\frac{c\dw{A,0}\,\kappa_0}{\Omega}
=
\frac{c\dw{A,0}}{\Omega\,R_0}
=
\frac{c\dw{A,0}}{\Omega\,r_0\,\cos \lambda_0}
\label{defeta}
\eeq
as the \emph{equilibrium parameter} of the flux tube, where $R_0$ is
the equilibrium radius of the flux ring in cylindrical coordinates, and
$r_0$ and $\lambda_0$ its radius and latitude, respectively, in
spherical coordinates.

\subsubsection{First-order perturbations}
The linear stability properties are determined through the first-order 
perturbation terms of Eq.\ (\ref{starteq}).
In the co-moving Frenet basis, the tangential, normal, and binormal
components are 
\beqa
\left( \frac{d}{d t} \vec{u} \right)_1 \cdot \vec{t}_0
- 
\left(
 \frac{1}{\rho}
 \frac{\partial}{\partial s}
 \frac{B^2}{8\pi}
\right)_1
+
\kappa_0
c\dw{A,0}^2
\left( \vec{n}_0 \cdot \vec{t}_1 \right)
-
2 \Omega \left( \vec{n}_0 \cdot \vec{u}_1 \right)
& = &
0
\label{eom:tan}
\\
\left( \frac{d}{d t} \vec{u} \right)_1 \cdot \vec{n}_0
-
\frac{\Delta\rho}{\rho}\bigg|_1
g_0 g_n
+
2 \Omega 
\left( \vec{t}_0 \cdot \vec{u}_1 \right)
+
\alpha
\left( \vec{u}\dw{rel,\perp,1} \cdot \vec{n}_0 \right)
\nonumber \\ {}
-
\kappa_0
c\dw{A,0}^2
\left(
 \frac{\kappa_1}{\kappa_0}
 +
 2\frac{B_1}{B_0}
 -
 \frac{ \rho_1}{\rho_0}
\right)
& = &
0
\label{eom:nor}
\\
\left( \frac{d}{d t} \vec{u} \right)_1 \cdot \vec{b}_0
-
\frac{\Delta\rho}{\rho}\bigg|_1
g_0 g_b
+
\kappa_0 c\dw{A,0}^2
\left( \vec{n}_0 \cdot \vec{b}_1 \right)
+
\alpha
\left( \vec{u}\dw{rel,\perp,1} \cdot \vec{b}_0 \right)
& = &
0
\ ,
\label{eom:bin}
\eeqa
with $g_{n}= \left( \vec{g}\cdot \vec{n}_0 \right)/g_0$ and $g_{b}=
\left( \vec{g}\cdot \vec{b}_0 \right)/g_0$.
For magnetic flux rings perpendicular to the stellar rotation axis it
is $\vec{t}_0= \vec{e}_\phi, \vec{n}_0= -\vec{e}_R$, and $\vec{b}_0=
\vec{e}_z$.
The first-order perturbations of all quantities (indicated by the index
`1') depend linearly on the Lagrangian displacement vector $\vec{\xi}=
\left( \xi_t, \xi_n, \xi_b \right)^T$, and are summarised in Appendix
\ref{pert}.
Their detailed derivation can be found, for example, in
\citet{1993GAFD...72..209, 1995GAFD...81..233} and
\citet{1998GApFD..89...75S}.

Except for the drag force, which comes from the perturbation of the
exterior, the velocity of the external plasma is taken to be unchanged
by the perturbation of the flux tube, that is, $\vec{u}\dw{e,1}= 0$.
The perturbations of the internal flow velocity and of the tube's axis
given by Eqs.\ (\ref{pert:v1}) and (\ref{pert:t1}), respectively, yield
the perturbation of the relative perpendicular flow velocity:
\beq
\vec{u}\dw{rel,\perp,1}
=
\vec{u}_1
-
\left( \vec{u}_1 \cdot \vec{t}_0 \right) \vec{t}_0
-
u_0 \vec{t}_1
=
\xi_{n,t} 
\vec{n}_0 
+
\xi_{b,t} 
\vec{b}_0 
\ .
\eeq
Derivatives with respect to time and arc length are indicated by the
indices $t$ and $s$, respectively.
At the bottom of stellar convection zones, the plasma pressure is
typically much higher than the magnetic pressure.
The stability analysis is therefore carried out in the approximation
$\beta= 8\pi p_0/B_0^2\gg 1$.
In the limit of an infinite radius of curvature and vanishing stellar
rotation (i.e.\ $\kappa\rightarrow 0, \Omega\rightarrow 0$), Eqs.\
(\ref{eom:tan})-(\ref{eom:bin}) reduce to the case of a horizontal flux
tube in a stratified atmosphere, which was investigated in
\citetalias{2007A&A...469...11H}.
Note that the present investigation disregards the possible influence
of enhanced inertia (factor $\mu$ in \citetalias{2007A&A...469...11H}),
owing to its conceptual difficulties in the case of curved thin flux
tubes \citep[cf.][and references therein]{1996A&A...312..317M}.

\subsubsection{Non-dimensional form}
The linearised equations of motion are transferred to non-dimensional
form by multiplying Eqs.\ (\ref{eom:tan})-(\ref{eom:bin}) with the
timescale $\tau= 1/\Omega$, which yields
\beq
 \tau^2
 \vec{\xi}_{,tt}
 +
 \mathcal{M}_{st}
 \frac{\tau}{\kappa_0}
 \vec{\xi}_{,st}
 +
 \mathcal{M}_{ss}
 \frac{1}{\kappa_0^2}
 \vec{\xi}_{,ss}
 +
\mathcal{M}_{t}
\tau
\vec{\xi}_{,t}
+
\mathcal{M}_{s}
\frac{1}{\kappa_0}
\vec{\xi}_{,s}
+
\mathcal{M}_{\xi}
\vec{\xi}
=
\vec{0}
\ .
\label{diffsys}
\eeq
The coefficient matrices are
\beqa
\mathcal{M}_{st}
& = &
2\,Ro\,\mathcal{E}
\label{coefma:st}
\\
\mathcal{M}_{ss}
& = &
-
2\,Ro\,\mathcal{E}
\label{coefma:ss}
\\
\mathcal{M}_{t}
& = &
2\,\left(Ro+1\right)\,\mathcal{Y}
+
\tilde{\alpha}\,\mathcal{X}
\label{coefma:t}
\\
\mathcal{M}_{s}
& = &
- 
2\,Ro\,\mathcal{Y}
+ 
\frac{\eta^2}{\gamma f_0}
\left( g\dw{n}\,\mathcal{Y} + g\dw{b}\,\mathcal{Z} \right)
\label{coefma:s}
\\
\mathcal{M}_{\xi}
& = &
\tilde{\omega}\dw{MBV}^2
\left(
\begin{array}{ccc}
0 & 0 & 0 \\
0 & g\dw{n} g\dw{n} & g\dw{n} g\dw{b} \\
0 & g\dw{n} g\dw{b} & g\dw{b} g\dw{b}
\end{array}
\right)
- 
\frac{\eta^2}{\gamma f_0}
\left(
\begin{array}{ccc}
0 & 0 & 0 \\
0 & 2 g\dw{n} & g\dw{b} \\
0 & g\dw{b} & 0
\end{array}
\right)
\ ,
\label{coefma:xi}
\eeqa
where 
\beqa
\mathcal{X}
=
\left(
\begin{array}{ccc}
0 & 0 & 0 \\
0 & 1 & 0 \\
0 & 0 & 1
\end{array}
\right)
, \quad
\mathcal{Y}
=
\left(
\begin{array}{ccc}
0 & -1 &  0 \\
1 & 0 & 0 \\
0 & 0 & 0
\end{array}
\right)
, \quad
\mathcal{Z}
=
\left(
\begin{array}{ccc}
0 & 0 & -1 \\
0 & 0 & 0 \\
1 & 0 & 0
\end{array}
\right)
\ ,
\eeqa
$\mathcal{E}$ is the unity matrix,
\beq
Ro
=
\frac{u_0 \kappa_0}{\Omega}
=
\sqrt{ 
 1 
 + 
 \eta^2
}
-
1
\label{defro}
\eeq
is the Rossby number,
\beq
f_0
=
H_p\,\kappa_0
\label{deff0}
\eeq
is the ratio between the local pressure scale height and the radius of
curvature, and $\tilde{\alpha}= \tau \alpha$ is the dimensionless
friction parameter.
The (dimensionless) magnetic Brunt-V\"ais\"al\"a frequency,
\beqa
\tilde{\omega}\dw{MBV}^2
& = &
\tau^2 \omega\dw{MBV}^2
=
-
\frac{\eta^2}{2 f_0^2} 
\left[
 \beta\,\delta
 -
 \frac{2}{\gamma}\,\left(\frac{1}{\gamma}-\frac{1}{2}\right)
\right]
\nonumber 
\\
& = &
\tau^2\,N^2
+
\frac{\eta^2}{\gamma^2 f_0^2} 
\left( 1 - \frac{\gamma}{2} \right)
\ ,
\label{defomegambv}
\eeqa
is a measure for axisymmetric buoyancy oscillations of magnetic flux
rings.
Owing to the stabilising effect of the magnetic field, its value is
higher than the non-magnetic Brunt-V\"ais\"al\"a frequency,
\beq
N^2
=
-
\frac{g}{H\dw{p}} \delta
\ ,
\label{defn}
\eeq
which depends on the superadiabaticity $\delta= \nabla - \nabla\dw{ad}$
and quantifies the stability of the stratification against convective
turnover.
Positive (negative) values of $\omega\dw{MBV}^2$ and $N^2$ indicate
stable (unstable) perturbations of the plasma in the presence and
absence of a magnetic field, respectively.

\subsubsection{Solar reference case}
For the numerical evaluation of linear stability properties, we shall
refer to a magnetic flux ring located in the middle of the solar
overshoot region.
At radius $r_0= 5.07\cdot 10^{10}\un{cm}$, the local density is
$\rho_0= 0.15\un{g/cm^3}$, the gravitation is $g_0=
5.06\cdot10^4\un{cm/s^2}$, the pressure scale heigh is $H\dw{p}=
5.52\cdot10^9\un{cm}$, and the superadiabaticity is $\delta_0=
-10^{-6}$.
The solar rotation rate is taken to be $\Omega_\odot=
2.69\cdot10^{-6}\un{s^{-1}}$ (i.e.\ $P\dw{rot}= 27\un{d}$) and the
ratio of specific heats $\gamma= 5/3$.
The local Brunt-V\"ais\"al\"a frequency is $N_0=
3\cdot10^{-6}\un{s^{-1}}$ and $\tilde{N}_0= N/\Omega= 1.11$.

\section{Linear stability analysis}
\label{linsa}

\subsection{Method based on sequence of determinants}
We consider perturbations in the form of discrete harmonic functions.
The exponential ansatz 
\beq
\vec{\xi}
=
\vec{\hat{\xi}}\,
\exp \left[ i\,\left(m\,\kappa\,s-\tilde{\omega}\,\Omega\,t\right) \right]
\label{expans}
\eeq
transforms Eq.\ (\ref{diffsys}) into the homogeneous algebraic system
\beqa
\left[
 \tilde{\omega}^2
 -
 m\,\tilde{\omega}\,
 \mathcal{M}_{st}
 +
 m^2\,
 \mathcal{M}_{ss}
 +
 i\,\tilde{\omega}\,
 \mathcal{M}_{t}
 -
 i\,m\,
 \mathcal{M}_{s}
 -
 \mathcal{M}_{\xi}
\right]\,
\vec{\hat{\xi}}
=
\vec{0}
\ ,
\label{algsys}
\eeqa
where $\tilde{\omega}= \tilde{\omega}\dw{r} + i \tilde{\omega}\dw{i}$
is the (dimensionless) complex eigenfrequency and $m$ is the azimuthal
wave number; owing to the $2\pi$-periodicity of the flux ring, the
latter is an integral number.
The determinant of the coefficient matrix of Eq.\ (\ref{algsys})
defines a 6th-degree dispersion polynomial in $\tilde{\omega}$, which
determines the stability properties of the magnetic flux ring:
eigenmodes with $\tilde{\omega}\dw{i}\le 0$ are (asymptotically)
stable, while those with $\tilde{\omega}\dw{i}> 0$ are unstable.
The coefficients of the dispersion polynomial are given in Appendix
\ref{coeffsgeneral}.

An explicit calculation of eigenfrequencies is not essential for the
determination of the stability criteria of a dynamical system, which
can be inferred from the coefficients of the dispersion polynomial by
using theorems from stability and control theory.
In our case, an instability implies a root of the dispersion equation
in the upper half of the complex plane.
We therefore use a generalised version of the Routh-Hurwitz theorem
\citep[Sect.\ 39]{marden1966}:
Assume the monic complex polynomial $f(z)= c_0 + c_1 z + \ldots +
c_{n-1} z^{n-1} + z^n\neq 0$ for $z$ real, and define the quantities
\beq
\Delta_{k}
=
\left( \frac{i}{2} \right)^k
D_k
\label{defdeltas}
\eeq
through the determinants $D_k$ of the $2k\times 2k$-principal minors
(i.e. the first $2k$ elements in the first $2k$ rows and columns) of
the square $2n\times 2n$-matrix\footnote{A bar over a complex quantity
denotes its complex-conjugate value.}
\beq
\mathcal{D}
=
\left(
\begin{array}{cccccccc}
1 & c_{n-1} & c_{n-2} & \ldots & c_{0} & 0 & 0 & \ldots \\
1 & \bar{c}_{n-1} & \bar{c}_{n-2} & \ldots & \bar{c}_{0} & 0 & 0 &
\ldots \\
0 & 1 & c_{n-1} & c_{n-2} & \ldots & c_{0} & 0 & \ldots \\
0 & 1 & \bar{c}_{n-1} & \bar{c}_{n-2} & \ldots & \bar{c}_{0} & 0 &
\ldots \\
\vdots & \vdots & \vdots & \vdots & \vdots & \vdots & \vdots & \vdots \\
\ldots & 0 & 1 & c_{n-1} & c_{n-2} & \ldots & c_{0} & 0 \\
\ldots & 0 & 1 & \bar{c}_{n-1} & \bar{c}_{n-2} & \ldots & \bar{c}_{0} &
0 \\
\ldots & 0 & 0 & 1 & c_{n-1} & c_{n-2} & \ldots & c_{0} \\
\ldots & 0 & 0 & 1 & \bar{c}_{n-1} & \bar{c}_{n-2} & \ldots & \bar{c}_{0} 
\end{array}
\right)
.
\label{matd}
\eeq
Then if $\Delta_k\neq 0$ for $k= 1,2,\ldots,n$, the number of zeros of
$f(z)$ in the upper half of the complex plane is equal to the number
sign changes in the sequence $\mathcal{V}= (1, \Delta_1, \Delta_2,
\ldots, \Delta_n)$.

The method is used in the following sections, which focus on special
cases of flux tube perturbations.

\subsection{Axisymmetric perturbations}
The flux tube equilibrium is indifferent with respect to axially
symmetric perturbations in the azimuthal direction.
Such perturbations are marginally stable and the associated
eigenfrequencies zero.
The remaining eigenfrequencies are determined by the quartic dispersion
polynomial 
\beq
\tilde{\omega}^4
+
2\,i\,\tilde{\alpha}\,\tilde{\omega}^3
+
\left( c_2 - \tilde{\alpha}^2 \right)\,\tilde{\omega}^2
+
i\,\tilde{\alpha}\,c_2\,\tilde{\omega}
+
c_0
=
0
\label{axipertdispeq}
\ ,
\eeq
with the real coefficients $c_2$ and $c_0$ given in Eqs.\
(\ref{c2tori}) and (\ref{c0tori}), respectively.
The elements of the $\mathcal{V}$-sequence are:
\beqa
\Delta_1
& = &
\frac{i}{2} D_1
=
\frac{i}{2} 
\left|
\begin{array}{cc}
1 & 2\,i\,\tilde{\alpha} \\ 1 & -2\,i\,\tilde{\alpha} 
\end{array}
\right|
=
2\,\tilde{\alpha}
\label{del1m0}
\\
\Delta_2
& = &
-
\frac{1}{4}
D_2
=
-
\frac{1}{4}
\left|
\begin{array}{cccc}
1 & 2\,i\,\tilde{\alpha} & c_{2}-\tilde{\alpha}^2 &
i\,\tilde{\alpha}\,c_2 \\
1 & -2\,i\,\tilde{\alpha} & c_{2}-\tilde{\alpha}^2 &
-i\,\tilde{\alpha}\,c_2 \\
0 & 1 & 2\,i\,\tilde{\alpha} & c_{2}-\tilde{\alpha}^2 \\
0 & 1 & -2\,i\,\tilde{\alpha} & c_{2}-\tilde{\alpha}^2
\end{array}
\right|
\nonumber
\\ 
& = &
-
2\,\tilde{\alpha}^2\,\left(c_2-2\,\tilde{\alpha}^2\right)
\label{del2m0}
\\
\Delta_3
& = &
-
\frac{i}{8}
D_3
\nonumber
\\
& = &
-
\frac{i}{8}
\left|
\begin{array}{cccccc}
1 & 2\,i\,\tilde{\alpha} & c_{2}-\tilde{\alpha}^2 & i\,\tilde{\alpha}\,c_2 & c_{0} & 0 \\
1 & -2\,i\,\tilde{\alpha} & c_{2}-\tilde{\alpha}^2 & -i\,\tilde{\alpha}\,c_2 & c_{0} & 0 \\
0 & 1 & 2\,i\,\tilde{\alpha} & c_{2}-\tilde{\alpha}^2 & i\,\tilde{\alpha}\,c_2 & c_{0} \\
0 & 1 & -2\,i\,\tilde{\alpha} & c_{2}-\tilde{\alpha}^2 & -i\,\tilde{\alpha}\,c_2 & c_{0} \\
0 & 0 & 1 & 2\,i\,\tilde{\alpha} & c_{2}-\tilde{\alpha}^2 & i\,\tilde{\alpha}\,c_2 \\
0 & 0 & 1 & -2\,i\,\tilde{\alpha} & c_{2}-\tilde{\alpha}^2 &
-i\,\tilde{\alpha}\,c_2 \\
\end{array}
\right|
\nonumber
\\
& = &
-
\tilde{\alpha}^3\,
\left(c_2-2\,\tilde{\alpha}^2\right)\,
\left(c_2^2-2\,\tilde{\alpha}^2\,c_2-4\,c_0\right)
\label{del3m0}
\\
\Delta_4
& = &
\frac{1}{16}
D_4
=
\frac{1}{16}
\left| \mathcal{D} \right|
\nonumber
\\
& = &
\tilde{\alpha}^4\,c_0\,
\left(c_2^2-2\,\tilde{\alpha}^2\,c_2-4\,c_0\right)^2
\label{del4m0}
\eeqa
According to $\Delta_2$, $\Delta_3$, and $\Delta_4$, instabilities
occur if $c_2> 2\,\tilde{\alpha}^2$ or $c_2^2-4c_0-2 \tilde{\alpha}^2
c_2< 0$ or $c_0< 0$, respectively.

The $\Delta_2$-criterion implies the condition
\beqa
\tilde{\omega}\dw{MBV}^2
& < &
\frac{2}{g_\perp^2}
\left( \frac{g\dw{n}}{\gamma f_0} - 2 \right) 
Ro \left( Ro + 2 \right)
-
\frac{4}{g_\perp^2}
-
\frac{2}{g_\perp^2}
\tilde{\alpha}^2
\label{m0inst:conv_ombv}
\ .
\eeqa
In the frictionless case, $\tilde{\alpha}= 0$, this criterion describes
a monotonic buoyancy-driven instability \citep{1995GAFD...81..233}.
Here, friction has a stabilising effect and reduces the critical magnetic 
Brunt-V\"ais\"al\"a frequency below which the instability sets in.

In our analysis of the $\Delta_3$-criterion, we follow the approach of
\citet{1995GAFD...81..233} and combine the coefficients $c_0$ and $c_2$
to eliminate $\tilde{\omega}\dw{MBV}^2$.
The resulting expression is used to eliminate $c_0$ from the
$\Delta_3$-criterion, which yields a quadratic expression in $c_2$.
If the discriminant,
\beqa
\lefteqn{
 \frac{d_{c_2}}{4}
 =
 \tilde{\alpha}^4
 -
 16 
 \frac{g\dw{b}^2}{g_\perp^2}
 \left(Ro+1\right)^2
 \tilde{\alpha}^2 
} \nonumber \\
& &
-
4 \frac{g\dw{b}^2}{g_\perp^2} 
\left[
 \frac{g_\perp}{\gamma f_0} Ro \left(Ro+2\right)
 -
 4 \frac{g\dw{n}}{g_\perp} \left(Ro+1\right)^2
\right]^2
\ ,
\label{discr}
\eeqa
of this quadratic expression is positive and the magnetic
Brunt-V\"ais\"al\"a frequency in the range
\beq
\tilde{\omega}\dw{c,-}^2
< 
\tilde{\omega}\dw{MBV}^2
<
\tilde{\omega}\dw{c,+}^2
\ ,
\label{m0inst:convosc_ombv}
\eeq
where
\beq
\tilde{\omega}\dw{c,\pm}^2
=
\frac{1}{g_\perp^2}
\left[
 \frac{8 g\dw{b}^2}{g_\perp^2} 
 -
 \left(
  4 
  -
  \frac{2 g\dw{n}}{\gamma f_0} 
  -
  \frac{8 g\dw{b}^2}{g_\perp^2} 
 \right)
 \eta^2
 -
 4
 -
 \tilde{\alpha}^2
 \pm
 \sqrt{\frac{d_{c_2}}{4}}
\right]
\ ,
\eeq
then $\Delta_3$ is negative.
In the frictionless case this criterion describes a buoyancy-driven
overstability which occurs if the environment is rotating
differentially \citep{1995GAFD...81..233}.

The $\Delta_4$-criterion applies to flux tubes outside the equatorial
plane only and implies
\beqa
\tilde{\omega}\dw{MBV}^2
& < &
\frac{1}{4}
\left( \frac{1}{\gamma f_0} \right)^2
\frac{\eta^4}{1+\eta^2}
\label{m0inst:posl_ombv}
\ .
\label{m0inst:posl_betadelta}
\eeqa
This instability is driven by the tension of magnetic field lines and
causes a monotonic movement of the flux ring to higher latitudes
\citep[cf.][]{1982A&A...106...58S}.
In contrast to the $\Delta_2$- and $\Delta_3$-criteria, the threshold
of this `poleward slip instability' depends neither on the friction
parameter nor on the local pressure scale height; the latter is
indicative for buoyancy-driven instabilities.

The poleward slip instability is the governing mechanism regarding
axially symmetric perturbations of flux rings outside the equatorial
plane.
This is shown by verifying that the sign change in the
$\mathcal{V}$-sequence first occurs between its last two elements.
At the $\Delta_4$-threshold, defined by equality of both sides of
relation (\ref{m0inst:posl_ombv}), that is $c_0= 0$, we have
\beqa
\left[ c_2-2\tilde{\alpha}^2\right]_{c_0=0}
& < &
\left[ c_2 \right]_{c_0=0}
\nonumber
\\
& = &
-
\left( \frac{1}{\gamma f_0} \right)^2\,
\frac{\eta^4}{4\,\left(Ro+1\right)^2}
+
2\,
\left(
 \frac{g\dw{n}}{\gamma f_0}
 -
 2
\right)
\eta^2
-
4
\nonumber
\\
& \le &
-
\left( \frac{1}{\gamma f_0} \right)^2\,
\frac{\eta^4}{4\,\left(Ro+1\right)^2}
+
2\,
\left(
 \frac{1}{\gamma f_0}
 -
 2
\right)
\eta^2
-
4
\nonumber
\\
& = &
-
\frac{ \left[ \frac{\eta^2}{\gamma f_0}
-4\,\left(1+\eta^2\right) \right]^2}{4\,\left(1+\eta^2\right)}
\le
0
\label{convisatc00}
\eeqa
and, by applying these relations, 
\beqa
\left[ c_2^2 - 4 c_0 - 2\tilde{\alpha}^2 c_2 \right]_{c_0= 0}
& = &
\left[ c_2^2 - 2\tilde{\alpha}^2 c_2 \right]_{c_0= 0}
\nonumber
\\
& = &
\left[ c_2 \left( c_2 - 2\tilde{\alpha}^2 \right) \right]_{c_0= 0}
\nonumber
\\
& = &
\left[ c_2 \right]_{c_0= 0}
\left[ c_2 - 2\tilde{\alpha}^2 \right]_{c_0= 0}
>
0
\ .
\eeqa
A magnetic flux ring which becomes subject to the poleward slip
instability is therefore still insusceptible to buoyancy-driven
instability mechanisms.
This applies to flux rings outside the equatorial plane only, where the
absolute term of the dispersion equation is not zero per se.

In the equatorial plane, buoyancy-driven instabilities occur if $c_2>
0$.
Since the discriminant $d_{c_2}= 4\tilde{\alpha}^2$ is positive, the
$\Delta_3$-criterion covers the range
\beq
2
\left(
 \frac{g\dw{n}}{\gamma f_0}
 -
 2
\right)
\eta^2
-
4
-
2 \tilde{\alpha}^2
< 
\tilde{\omega}\dw{MBV}^2
<
2
\left(
 \frac{g\dw{n}}{\gamma f_0}
 -
 2
\right)
\eta^2
-
4
\label{m0inst:conveqplane_ombv}
\ ,
\eeq
and therefore extends the instability criterion (\ref{m0inst:conv_ombv}) 
to higher magnetic Brunt-V\"ais\"al\"a frequencies.
In analogy with the results of \citet{1995GAFD...81..233} for the
frictionless case, we expect that in the presence of differential
rotation the $\Delta_3$-criterion describes a buoyancy-driven
overstability.
In the present case of solid body rotation, explicit calculations of
the eigenfrequencies show that (within computational accuracy) axially
symmetric instabilities are monotonic.

\begin{figure}
\includegraphics[width=\hsize]{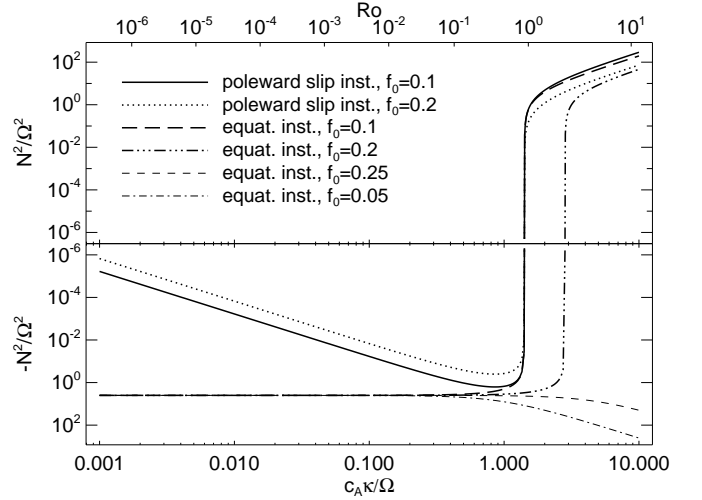}
\caption{Thresholds of the poleward slip instability and axially
symmetric equatorial instability for different length scale ratios
$f_0= H\dw{p}\,\kappa_0$.
Flux tube equilibria with parameters located above (below) a curve are
insusceptible (susceptible) to the respective instability.}
\label{axipertcomp}
\end{figure}

In the determination of the critical equilibrium parameter, beyond
which magnetic flux rings are subject to unstable axially symmetric
perturbations, the dependence of the magnetic Brunt-V\"ais\"al\"a
frequency on the field strength must be taken into account.
To this end, we use Eq.\ (\ref{defomegambv}) to substitute
$\tilde{\omega}\dw{MBV}^2$ in Eqs.\ (\ref{m0inst:posl_ombv}) and
(\ref{m0inst:conveqplane_ombv}).
The equation for the threshold of the poleward slip instability is
\beq
\left(3-2\gamma\right)\,\left( \frac{1}{\gamma f_0} \right)^2
\eta^4
+
\left[
 4 \tilde{N}^2
 +
 \left(4-2\gamma\right)
 \left( \frac{1}{\gamma f_0} \right)^2
\right]
\eta^2
+
4 \tilde{N}^2
=
0
\ .
\label{polslipeq}
\eeq
This instability occurs in a stably stratified environment (i.e.\
$\tilde{N}^2> 0$) if the equilibrium parameter is sufficiently large
(Fig.\ \ref{axipertcomp}).
In contrast, a flux ring located in a convectively unstable environment
is insusceptible to this instability mechanism if the local
Brunt-V\"ais\"al\"a frequency is higher than
\beq
\tilde{N}\dw{min}^2
=
\frac{\sqrt{2\gamma-3} - \gamma + 1}{2\gamma^2 f_0^2}
\simeq
-1.6\cdot 10^{-2}/f_0^2
\label{defbw2min}
\eeq
and the equilibrium parameter is within a range of values delimited by
the solutions of Eq.\ (\ref{polslipeq}); an upper boundary of this
range is
\beq
\eta (\tilde{N}^2=0)
=
\sqrt{\frac{4-2\gamma}{2\gamma-3}}
=
\sqrt{2}
\ .
\label{defetat}
\eeq
The threshold of the buoyancy-driven instability in the equatorial plane is
determined by 
\beqa
\left[ 
 \left(1-\frac{\gamma}{2}\right)
 \left( \frac{g\dw{n}}{\gamma f_0} \right)^2
 -
 2 \frac{g\dw{n}}{\gamma f_0} 
 +
 4
\right]
\eta^2
+
g\dw{n}^2 \tilde{N}^2 +4
=
0
\ .
\eeqa
In a stably stratified environment a flux ring is susceptible to this
instability mechanism provided that the equilibrium parameter is
sufficiently large and the ratio between pressure scale height and
radius of curvature is in the range
\beq
g\dw{n}
\frac{1-\sqrt{2\,\gamma-3}}{4\,\gamma}
<
f_0
<
g\dw{n}
\frac{1+\sqrt{2\,\gamma-3}}{4\,\gamma}
\ .
\label{f0range}
\eeq
If the length scale ratio is outside this range, the axisymmetric
equatorial instability can only occur in a convectively unstable
environment (Fig.\ \ref{axipertcomp}, thin lines).
With $\gamma= 5/3$ and $g\dw{n}= 1$, the range is $0.063< f_0< 0.24$.
In the overshoot region of the Sun it is (at the equator) $f_0\approx
0.1$.

Both poleward slip and equatorial instabilities are not caused by
friction in the first place and occur in the frictionless case as
well.
There are no specific friction-induced instabilities regarding axially
symmetric perturbations of magnetic flux rings, though friction may
affect the occurrence of overstabilities in the presence of
differential rotation.
The absence of friction-induced instabilities is consistent with our
finding in \citetalias{2007A&A...469...11H} that their occurrence
depends on the relation between flow and phase velocity.
Since axially symmetric perturbations do not yield propagating wave
signals the method of principal minors recovers the (modified)
instability mechanisms which also occur in the frictionless case.

\subsection{Flux tubes in the equatorial plane}
We now consider flux rings in the equatorial plane and derive
analytical criteria for friction-induced instabilities with azimuthal
wave numbers $m\ge 1$.
In the equatorial plane binormal perturbations are decoupled from
perturbations within the osculating plane, so that the dispersion
equation factorises into a quadratic and a quartic polynomial.

The eigenfrequencies of binormal perturbations, 
\beq
\tilde{\omega}
=
m\,Ro
-
i\,\frac{\tilde{\alpha}}{2}
\pm
\sqrt{
 m^2\,\eta^2
 -
 i\,\tilde{\alpha}\,m\,Ro
 -
 \frac{\tilde{\alpha}^2}{4}
}
\ ,
\label{binoomti}
\eeq
reduce in the frictionless case to $m\left( Ro \pm \eta \right)$.
The corresponding phase velocities $v\dw{ph}= u_0 \pm c\dw{A}$ imply
perturbations which propagate with the Alfv\'en velocity in the
prograde and the retrograde direction relative to the internal plasma
flow.
In the frictional case, a binormal eigenmode is unstable if
\beqa
\Im
\left(
 \sqrt{
  m^2\,\eta^2
  -
  i\,\tilde{\alpha}\,m\,Ro
  -
  \frac{\tilde{\alpha}^2}{4}
 }
\right)
> 
\frac{\tilde{\alpha}}{2}
\ ,
\eeqa
which requires flow velocities $u_0> c\dw{A}$.
This requirement is never fulfilled, because the mechanical equilibrium 
permits sub-Alfv\'enic flow velocities only.
Binormal perturbations of flux rings in the equatorial plane are thus
stable both in the presence and in the absence of friction.

The eigenfrequencies of perturbations in the osculating plane are
determined by a monic quartic polynomial, whose coefficients are given
by Eqs.\ (\ref{c3equa})-(\ref{c0equa}) in Appendix \ref{coeffsequa}.
The associated $\mathcal{V}$-sequence is 
\beqa
\Delta_1
& = &
\tilde{\alpha}
\label{del1equa}
\\
\Delta_2
& = &
\tilde{\alpha}^2\,
\left[
 4\,\left(1+\eta^2\right)
 +
 2\,m^2\,Ro
 -
 2\,\frac{g\dw{n}}{\gamma f_0}\,\eta^2
 +
 g\dw{n}^2\,\tilde{\omega}\dw{MBV}^2
\right]
\label{del2equa}
\\
\Delta_3
& = &
\tilde{\alpha}^3\,m^2\,\eta^2\,
\left[
 \left(2\,m^2\,Ro+g\dw{n}^2\,\tilde{\omega}\dw{MBV}^2\right)\,F_1
 -
 F_2
\right]
\label{del3equa}
\\
\Delta_4
& = &
\tilde{\alpha}^4\,m^6\,\eta^4\,
\left[
 \left( \frac{g\dw{n}}{\gamma f_0} \right)^2\,\eta^2
 -
 4\,\frac{g\dw{n}}{\gamma f_0}\,\eta^2
 -
 4
\right]^2\,
\nonumber \\ & &
\times
\left[
 4\,\left(m^2-1\right)\,Ro^2
 -
 \left( \frac{g\dw{n}}{\gamma f_0} \right)^2\,\eta^4
 +
 2\,g\dw{n}^2\,\tilde{\omega}\dw{MBV}^2\,Ro
\right]
\label{del4equa}
\eeqa
with
\beqa
F_1
& = &
4 \left(4 \eta^2 - 2 Ro + 1\right)
+
4 \frac{g\dw{n}}{\gamma f_0} \left(Ro-2\eta^2\right)
+
\left(\frac{g\dw{n}}{\gamma f_0}\right)^2\eta^2
\label{deff1}
\\
F_2
& = &
16 \left(1+\eta^2\right) \left(2 Ro-1\right)
-
8 \frac{g\dw{n}}{\gamma f_0} 
Ro \left(4 \eta^2+3 Ro+8\right)
\nonumber \\ & & {}
-
4\left( \frac{g\dw{n}}{\gamma f_0} \right)^2
\eta^2 \left(Ro^2-3\right)
+
2\left( \frac{g\dw{n}}{\gamma f_0} \right)^3\eta^4
\label{deff2}
\eeqa
In Eqs.\ (\ref{del1equa})-(\ref{del4equa}), the friction parameter
occurs as a multiplicative factor only: $\Delta_k\propto
\tilde{\alpha}^k$.
The instability thresholds therefore depend on the presence of friction
only, but not on its strength.

It is sufficient to analyse the lowest-order eigenmode, because
$\Delta_2, \Delta_3$, and $\Delta_4$ increase with the azimuthal wave
number:
\beq
\frac{d\Delta_k}{dm}> 0
\quad \Rightarrow \quad
\Delta_k (m_1) < \Delta_k (m_2) 
\quad \textrm{for} \quad 0< m_1< m_2
\label{delminc}
\eeq
The increase of $\Delta_3$ with $m$ is not obvious, since it depends on 
$F_1$. 
Regarding $F_1$ as a quadratic function in $g\dw{n}$, for example, its
roots $g\dw{n,\pm}$ are complex-conjugate, so that for real-valued
$g\dw{n}$ the sign of $F_1$ is unique.
Since the function is positive for, say, $g_n= 0$, it follows $F_1>
0$.

An instability is indicated by $\Delta_2< 0$, that is,
\beqa
\tilde{\omega}\dw{MBV}^2
& < &
\frac{2\,\left(Ro+2\right)}{g\dw{n}^2}
\left[ \left(\frac{g\dw{n}}{\gamma f_0}-2\right)\,Ro-1\right]
\ ,
\eeqa
or by $\Delta_3< 0$, implying 
\beqa
\tilde{\omega}\dw{MBV}^2
& < &
\frac{F_2-2\,Ro\,F_1}{g\dw{n}^2\,F_1}
\ .
\eeqa
The $\Delta_4$-instability occurs for
\beqa
\tilde{\omega}\dw{MBV}^2
& < & 
\frac{1}{2}
\left( \frac{1}{\gamma f_0} \right)^2
Ro\,\left(Ro+2\right)^2
\label{iscoscu:ombv}
\ .
\eeqa
We shall show that the first sign change in the sequence $\mathcal{V}$ 
occurs at its end, so that at the threshold of the
$\Delta_4$-instability both $\Delta_2$ and $\Delta_3$ are still
positive; from $\tilde{\alpha}> 0$ follows $\Delta_1> 0$.
After substitution of $\tilde{\omega}\dw{MBV}^2$ in Eq.\
(\ref{del2equa}) with the right side of Eq.\ (\ref{iscoscu:ombv}) we
get
\beq
\left[ \Delta_2 \right]_{c_0=0}
=
\frac{\tilde{\alpha}^2}{2}\left(Ro+2\right)
\left[
 \left( \frac{g\dw{n}}{f_0\,\gamma} \right)^2\,\eta^2
 +
 4\,\left(2-\frac{g\dw{n}}{\gamma f_0}\right)\,Ro
 +
 4
\right]
\ .
\eeq
Following the same argumentation as after Eq.\ (\ref{delminc}) for
function $F_1$, it can be shown that the term in square brackets is
positive.
Therefore, $\Delta_2> 0$.
For $\Delta_3$ we find
\beqa
\left[ \Delta_3 \right]_{c_0=0}
=
\frac{\tilde{\alpha}^3}{2}
\eta^2\left(Ro+2\right)
\left[
 \left( \frac{g\dw{n}}{\gamma f_0} \right)^2\,\eta^2
 -
 4\,\left( \frac{g\dw{n}}{\gamma f_0} \right)\,\eta^2
 -
 4
\right]^2
\ ,
\eeqa
which is non-negative.
Relation (\ref{iscoscu:ombv}) is therefore the instability criterion for 
non-axial symmetric perturbations of magnetic flux rings in the
equatorial plane.

\begin{figure}
\includegraphics[width=\hsize]{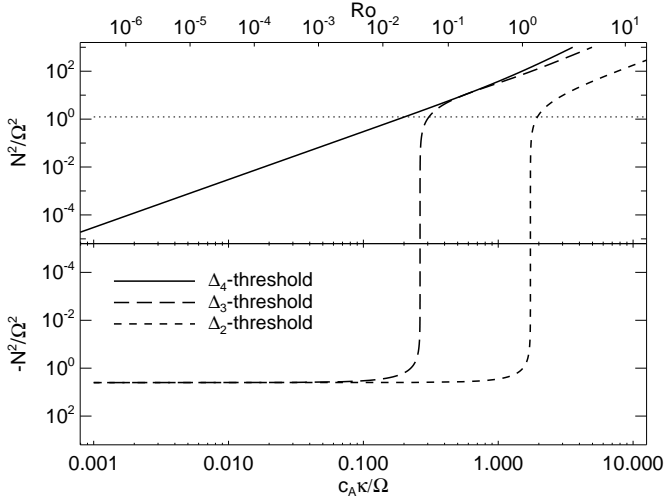}
\caption{Thresholds of the friction-induced instability (\emph{solid
line}) and of buoyancy-driven instabilities (\emph{broken lines}) for 
non-axial symmetric perturbations (for $f_0= 0.1$).
The dotted, horizontal line indicates the stratification parameter
$\tilde{N}^2= 1.23$ in the middle of the solar overshoot region.}
\label{critmag3}
\end{figure}

Taking the field dependence of the magnetic Brunt-V\"ais\"al\"a
frequency into account, the threshold of the $\Delta_4$-instability is
given by
\beq
\tilde{N}^2
=
\frac{1}{2}
{\left( \frac{1}{\gamma f_0} \right)^2 Ro
 \left(Ro+2\right)
 \left(Ro+\gamma\right)
}
\ .
\label{iscfieq}
\eeq
Its dependence on the equilibrium parameter is shown in Fig.\
\ref{critmag3}, together with the thresholds of the $\Delta_2$- and
$\Delta_3$-criteria.
In the equatorial plane, magnetic flux rings are unstable against
non-axial symmetric perturbations if they are located in a convectively
unstable stratification.
If the environment is stably stratified, the instability occurs
provided the equilibrium parameter is sufficiently large.

Since there is only one sign change in the $\mathcal{V}$-sequence, only
one eigenmode is unstable yielding growing perturbations.
A comparison with Eq.\ (\ref{defomegambv}) shows that Eq.\
(\ref{iscfieq}) is independent of the pressure scale height, so that
the underlying instability is not caused by buoyancy.
It does not exist in the frictionless case and is therefore specifically 
induced by the frictional interaction between the magnetic flux tube
and its environment.
The driving mechanism of the friction-induced instability has been
described in detail in \citetalias{2007A&A...469...11H}.

\begin{figure}
\includegraphics[width=\hsize]{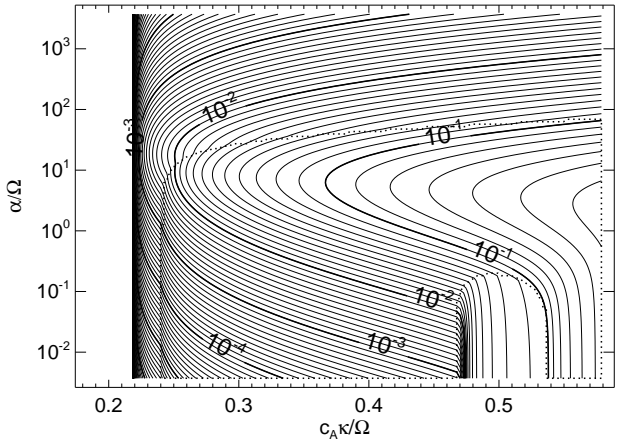}
\caption{Growth rates, $\Im(\omega)/\Omega$, of magnetic flux tubes in
the equatorial plane.
The threshold to the left is determined by the friction-induced
instability with azimuthal wave number $m= 1$.
In the region enclosed by dotted lines, $m= 2$ modes have the highest
growth rates.
The dimensionless Brunt-V\"ais\"al\"a frequency is $N^2/\Omega^2= 1.23$
(cf.\ Fig.\ref{critmag3}) and $f_0= 0.109$.}
\label{quartic}
\end{figure}

The basic condition which discriminates between stable and unstable
flux tube equilibria is the $\Delta_4$-criterion with azimuthal wave
number $m= 1$.
Beyond that threshold, higher-order eigenmodes can be unstable with growth 
rates higher than those of the $m=1$ mode.
The determination of the fastest growing perturbations requires the
explicit calculation of all eigenfrequencies.
Keeping the equilibrium parameter constant, growth rates decrease 
if friction is strong (Fig.\ \ref{quartic}).
This is caused by the dissipative nature of friction, which opposes
relative plasma motions perpendicular to the tube axis.
If friction is weak, the dissipative effect is outbalanced by the
energy transfer from the relative motion of internal and external
plasma through frictional coupling.
Further to friction, the growth times of unstable perturbations still 
depend on magnetic buoyancy and tension.
Growing perturbations of flux tubes which in the frictionless case are
subject to Parker-type instabilities experience in the frictional case
still a strong magnetic buoyancy which, gives rise to relative high
growth rates (cf.\ Fig.\ \ref{quartic}, lower-right corner).

An important aspect of our approach using the principal minors of the
coefficient matrix $\mathcal{D}$ is that, for a polynomial of degree
$n$ with $\Im \left( c_0 \right)= 0$, the last $\Delta_k$ in the
sequence $\mathcal{V}$ depends linearly on the absolute term of the
dispersion polynomial: $\Delta_n\propto c_0$.
This can be seen by comparing Eq.\ (\ref{del4equa}) with Eq.\
(\ref{c0equa}) and also in Eq.\ (\ref{del4m0}).
A sign change of the absolute term of the dispersion polynomial thus
indicates a friction-induced instability with, at least, one
eigenfrequency being zero at the threshold $c_0= 0$.
This result agrees with our finding in
\citetalias{2007A&A...469...11H}, that the onset of a friction-induced
instability is characterised by the reversal of the propagation
direction of a wave mode from retrograde, with $\Re \left(
\tilde{\omega} \right)< 0$, to prograde, with $\Re \left(
\tilde{\omega} \right)> 0$.

\subsection{General case}
\label{generalcase}
In case of magnetic flux rings parallel to but outside the equatorial
plane, gravitation causes a coupling of perturbations parallel and
perpendicular to the rotation axis.
The eigenfrequencies are determined by a 6th-degree dispersion
polynomial, whose coefficients are given in Appendix
\ref{coeffsgeneral}.
The method based on the principal minors is still applicable, but the
analytical expressions for $\Delta_k$ are cumbersome and not of much
practical use.
Guided by our previous results, we take the reversal of the propagation
direction of an eigenmode as the criterion for the friction-induced
instability, implying\footnote{The minus sign is caused by the factor
$(i/2)^6$ in Eq.\ (\ref{defdeltas}).} $\Delta_6\propto -c_0< 0$.
The resulting instability criterion is formally identical with
condition (\ref{iscoscu:ombv}).

\begin{figure}
\includegraphics[width=\hsize]{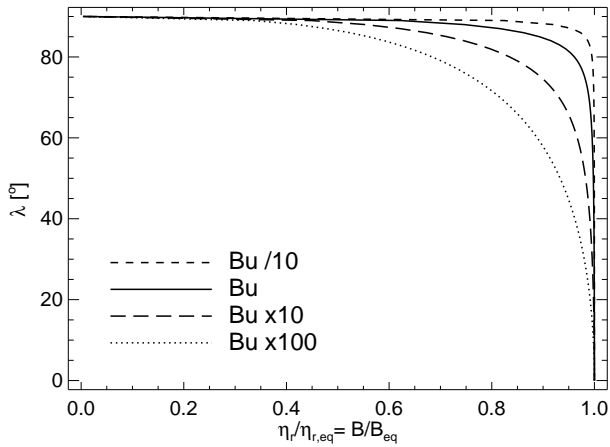}
\caption{Latitude dependence of the friction-induced instability
threshold for different Burger numbers; in the solar reference case, it
is $Bu= (N H\dw{p}/\Omega r_0)^2= 1.47\cdot10^{-2}$.
Note that the equatorial value $\eta\dw{r,eq}$ increases with the
Brunt-V\"ais\"al\"a frequency of the stratification (i.e. for high
$Bu$).}
\label{latidep2}
\end{figure}

The latitude-dependence of the instability threshold, Eq.
(\ref{iscfieq}), is mediated through the dependence of Rossby number
and equilibrium parameter on the curvature of the equilibrium flux
tube.
Consider, for example, similar magnetic flux rings at a given depth of
a spherically symmetric star.
Since the Brunt-V\"ais\"al\"a frequency is the same for each flux tube
at different latitudes, the critical flow velocity decreases with
increasing latitude: $u\dw{0,crit}\propto \cos \lambda_0$.
Owing to the mechanical equilibrium condition (\ref{defuequi}), this
implies that at higher latitudes the friction-induced instability sets
in at lower field strengths (Fig.\ \ref{latidep2}).
If the latitude dependence of all parameters is taken into account, the
threshold is determined through the bi-cubic polynomial
\beqa
\lefteqn{
 \eta_r^6
 -
 \left(\gamma-2\right) \gamma \cos^2 \lambda
 \eta_r^4
 +
 4 \left(\gamma-1\right) \gamma^2 Bu \cos^2\lambda 
 \eta_r^2
} 
\nonumber \\
& &
{} -
4 \gamma^4 Bu^2 \cos^2\lambda 
=
0
\ .
\label{critetalat}
\eeqa
The index `$r$' of the equilibrium parameter $\eta_r= c\dw{A,0}/\Omega
r_0$ indicates that this quantity is a function of the equilibrium
depth, $r_0$, since all latitude dependencies are now explicitely taken
into account through the cosine-terms in the coefficients.
For $\lambda= 0$, Eq.\ (\ref{critetalat}) is equivalent to Eq.\
(\ref{iscfieq}).
The Burger number $Bu= (N H\dw{p}/\Omega r_0)^2$ is a measure for the
ratio of buoyancy and Coriolis force, with low (high) values indicating
dynamics dominated by rotation (stratification).
Fig.\ \ref{latidep2} shows a comparison of the instability threshold
for different Burger numbers, corresponding to, for example, different
stellar rotation rates or Brunt-V\"ais\"al\"a frequencies.

\begin{figure*}
\includegraphics[width=\hsize]{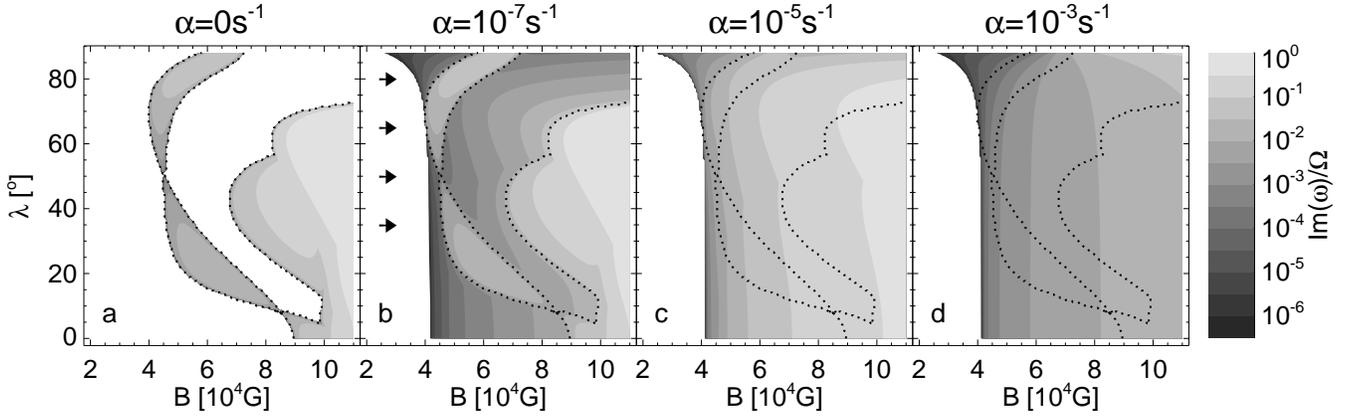}
\caption{Growth rates of Parker-type and friction-induced instabilities
of magnetic flux rings in the middle of the solar overshoot region.
Shaded areas indicate unstable equilibria, with dotted lines marking
boundaries of Parker-type instabilities in the frictionless case
(\emph{panel a}).
Growth rates first increase with friction (\emph{panel b}), then reach
a maximum (\emph{panel c}), and decrease for very strong friction (panel
d).
Arrows in panel b indicate the latitudes shown in Fig.\
\ref{sexticthres}.}
\label{sexticsun}
\end{figure*}

\begin{figure}
\includegraphics[width=\hsize]{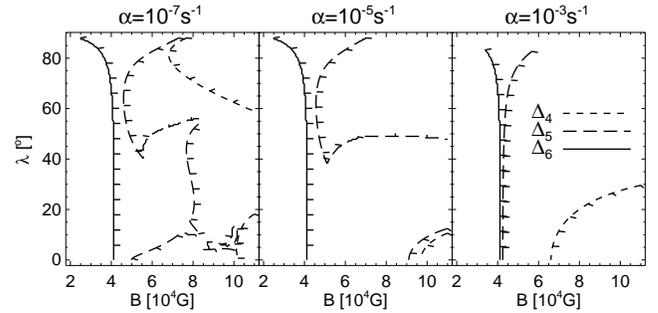}
\caption{Sign changes of $\Delta_{4}$, $\Delta_5$, and $\Delta_6$ in
the solar reference case, corresponding to the instability diagrams in
Fig.\ \ref{sexticsun}.
Tickmarks along each line indicate the downhill direction of the
$\Delta_k$-functions.
Within the shown range of field strengths and latitudes, $\Delta_{1-3}>
0$.}
\label{sexticsundeltas}
\end{figure}

\begin{figure}
\includegraphics[width=\hsize]{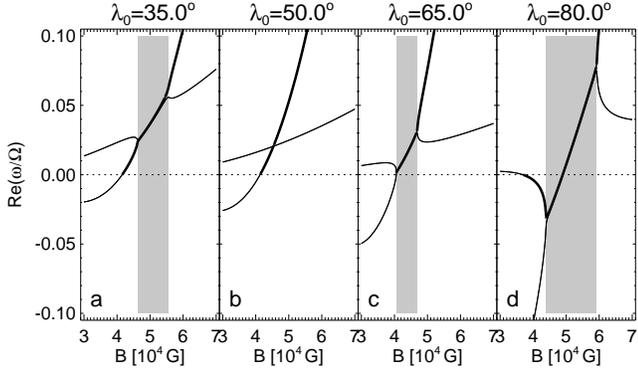}
\caption{Oscillation frequencies of slow eigenmodes near to the threshold
of friction-induced instabilities in the solar reference case.
Each panel corresponds to a latitudinal cut in the stability diagram
Fig.\ \ref{sexticsun}, panel b; the friction parameter is
$\alpha= 10^{-7}\un{s^{-1}}$.
Thick lines indicate an unstable eigenmode. 
In the gray shaded regions the instability is predominantly driven by
magnetic buoyancy, outside by frictional coupling.
Since the azimuthal wave number is $m= 1$, we have $Re(\omega/\Omega)=
v\dw{ph}/v\dw{rot}$.}
\label{sexticthres}
\end{figure}

We numerically investigate the friction-induced instability properties
of flux rings outside the equatorial plane on the basis of the solar
reference case (Fig.\ \ref{sexticsun}).
The calculation of the $\mathcal{V}$-sequence confirms that the
$\Delta_6$-criterion determines the onset of friction-induced
instabilities (Fig.\ \ref{sexticsundeltas}).
For field strengths beyond this threshold, the sign change in the
$\mathcal{V}$-sequence can occur at $\Delta_5$ or $\Delta_4$, but in
the parameter range considered here there is always only one sign
change, indicating a single unstable eigenmode.
Given that the friction-induced instability sets in once an eigenmode
reverses its propagation direction, the retrograde and prograde slow
wave modes are liable to become unstable since the moduli of their
phase velocities are smallest.
Owing to the prograde plasma flow inside the flux tube, the retrograde
wave is advected against its propagation direction and (the modulus of)
its phase velocity closest to zero.
In most cases it is the retrograde eigenmode which becomes frictionally
unstable once its propagation reverses into prograde direction.
An exception to this rule occurs if the threshold of the
friction-induced instability coincides with strong driving caused by
magnetic buoyancy.
In the frictionless case, Parker-type instabilities are characterised
by a merging of two eigenmodes: the phase velocities of both modes are
identical and the growth rates of opposite sign (cf.\ Fig.\
\ref{sexticthres}, panel a).
If the merging coincides with the threshold of the friction-induced
instability, the phase velocities of both eigenmodes are zero (panel
c).
In that situation, the prograde eigenmode can reverse its propagation
to the retrograde direction and become frictionally unstable (panel d).

In contrast to the frictionless case, where magnetic flux rings at high
latitudes may be stable up to high field strengths, flux rings subject
to friction are generally unstable if their field strength is beyond
the threshold given by Eq.\ (\ref{iscfieq}).
The growth rates depend on the dominating driving mechanisms, which in
the parameter range considered here are frictional coupling and
magnetic buoyancy; tension-driven poleward slip instabilities typically
occur at higher field strengths.
If friction is weak, growth rates are significant only if the flux tube
equilibrium is in a parameter domain of Parker-type instabilities
(Fig.\ \ref{sexticsun}, panel b).
If friction is strong, a differentiation between Parker-stable and
Parker-unstable domains becomes irrelevant, since the growth times are
comparable.
In case of very strong friction, growth rates decrease as flux tube
motions are more and more harnessed by the environment.

\section{Discussion}
\label{disc}
The criteria for friction-induced instability of straight and toroidal
magnetic flux tubes are identical.
From the equilibrium condition, Eq.\ (\ref{defuequi}) follows
\beq
Ro
=
\frac{2 M\dw{A}^2}{1-M\dw{A}^2}
\ ,
\eeq
where $M\dw{A}= u_0/c\dw{A}$ is the Alfv\'enic Mach number, which is
limited in the toroidal case to values $M\dw{A}< 1$.
Substitution of the Rossby number in the instability criterion, Eq.\
(\ref{iscoscu:ombv}), yields
\beq
\omega\dw{MBV}^2
< 
\left( \frac{c\dw{A}}{\gamma H\dw{p}} \right)^2
\frac{1}{1 - M\dw{A}^2}
\ ,
\label{ombvmalf}
\eeq 
which is the instability criterion of straight, horizontal flux tubes
in the limit of long wavelengths \citepalias[cf.\ Eq.\ (32)
in][]{2007A&A...469...11H}.
In the limit of an infinite radius of curvature Eq.\ (\ref{ombvmalf})
becomes
\beq
\omega\dw{MBV}^2< \left(\frac{c\dw{A}}{\gamma H\dw{p}} \right)^2
\quad \rightarrow \quad
\beta
\delta
>
-
\frac{1}{\gamma}
\ ,
\label{critSvB}
\eeq
since $u_0\rightarrow 0$ for $\kappa_0\rightarrow 0$.
Relation (\ref{critSvB}) is the instability criterion of a straight
flux tube in a stratified environment without a velocity difference
between internal and external plasma \citep{1982A&A...106...58S}.

In contrast to the case of straight flux tubes, in which the flow
velocity is a free parameter, the relative flow velocity inside
toroidal flux tubes is fixed by the equilibrium condition.
This implies an explicit dependence of the friction-induced instability
on the curvature and field strength of the flux ring.
The instability criterion, which relates the flow velocity with the
phase velocity of a perturbation, thus determines a critical magnetic
field strength (or, more general, Alfv\'en velocity) beyond which the
friction-induced instability sets in.
This critical field strength is lower than the threshold of the
buoyancy-driven instability.
A preliminary numerical parameter study indicates that this property
applies to a broad range of stellar rotation rates and equilibrium
depths in the overshoot region.
From this, we conjecture that it is the friction-induced instability
which determines whether flux ring equilibria in the overshoot region
are stable or not.

\begin{figure*}
\includegraphics[width=\hsize]{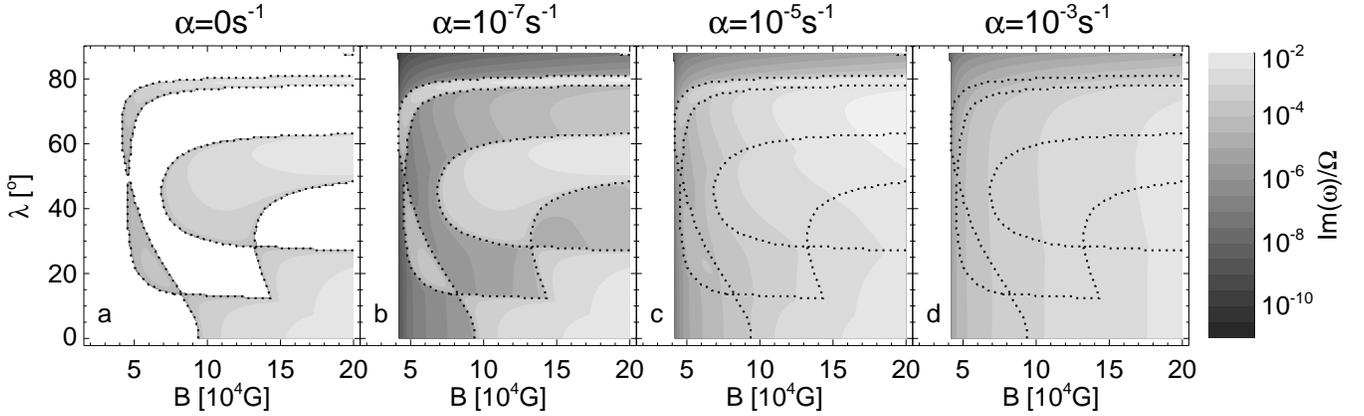}
\caption{Like Fig.\ \ref{sexticsun}, but for a rapidly rotating star with 
a rotation period of $P\dw{rot}= 2\un{d}$. 
The Burger number is $Bu= 8.04\cdot 10^{-5}$.
In accordance with the results shown in Fig.\ \ref{latidep2}, the
latitude dependence of the threshold is weaker than in the solar
reference case and only effective at very high latitudes.}
\label{sexticfast}
\end{figure*}

Virtually all flux tube equilibria beyond the instability threshold are
unstable.
This is important, for example, in the case of fast stellar rotation.
Stability analyses of flux rings in rapidly rotating stars
\citep[e.g.][]{1995GAFD...81..233, 2003A&A...405..291H} show that, in
the frictionless case, for high magnetic field strengths isolated
regions of stable equilibria exist, in which flux tubes are not subject
to the Parker-type instability (see Fig.\ \ref{sexticfast}, panel a).
These field strengths are well beyond the threshold of the 
friction-induced instability.
Therefore, if friction is taken into account, these equilibria are
unstable with high growth rates (Fig.\ \ref{sexticfast}, panels c and
d).

The possibility to store magnetic flux in the overshoot region depends
on the growth rate of the instabilities.
An evaluation of the growth rates of friction-induced instabilities
requires an estimate for the friction parameter $\alpha$.
A comparison of drag force, Eq.\ (\ref{defdrag}), and friction force,
Eq.\ (\ref{frictrans}), yields
\beq
\alpha
\sim 
\frac{C\dw{D}}{\pi a} \left| \vec{u}\dw{rel,\perp} \right| 
\ .
\eeq
The friction parameter of a magnetic flux tube which gives rise to a
small sunspot is estimated through non-linear numerical simulations
\citep[see][for a description of the code]{1995ApJ...441..886C}.
With $B_0= 10^5\un{G}, a_0= 1000\un{km}$ (i.e., $\Phi\simeq
3\cdot10^{21}\un{Mx}$) and $C\dw{D}= 1$ we obtain $\alpha\sim
10^{-7}\un{s^{-1}}$.
The results of the stability analysis in Fig.\ \ref{sexticsun} show
that for this value of the friction parameter growth rates are several
orders of magnitude lower than the rotation rate.
Sunspot-producing flux tubes are thus hardly susceptible to 
friction-induced instabilities.

Thin magnetic flux tubes are weeded out of the overshoot region on
shorter time scales than thick flux tubes, and should therefore occur
more often.
This agrees with the trend indicated by umbral areas observed on the
Sun, which follow a log-normal distribution with small surface features
being more numerous than large features
\citep[e.g.][]{2005A&A...443.1061B}.
If the higher number of small surface features is a consequence of
higher friction-induced emergence rates, the dynamics and stability
properties of thin magnetic flux tubes are dominated by friction.
In that case, the latitudinal distribution of surface features may also
depend on the size of the magnetic feature, since thinner flux tubes
experience a stronger deflection to higher latitudes by meridional
circulation \citep[e.g.][]{2006MNRAS.369.1703H}.
The influence of meridional circulation would possibly be diluted by
the action of convective motions on rising flux tubes.
But since the former effect is systematic and the latter random, a
discernable signature may persist in the latitudinal distribution.
Even if thin flux tubes do not make it to the surface due to convective
interaction, their susceptibility to friction-induced instabilities
implies a decrease of magnetic flux in the overshoot region, which
affects the budget and amplification of magnetic fields in the
tachocline region.

The numerical results shown in Sect.\ \ref{generalcase} are based on
magnetic flux rings located in the middle of the solar overshoot layer.
Owing to the strong dependence of stability properties on the
superadiabaticity, flux tubes located deeper in the overshoot region
require significantly higher field strength to become buoyancy
unstable.
The friction-induced instability eases the need for very high field
strengths, since its threshold is below the critical field strength of
the buoyancy-driven instability.
For example, a flux ring located in the equatorial plane close the
radiative core, in an environment with $\delta= -10^{-4}$ ($N=
3\cdot10^{-5}\un{s^{-1}}$), becomes frictionally unstable for field
strengths above $3.5\cdot10^5\un{G}$, which is well below the critical
field strength of the Parker-type instability at
$6.2\cdot10^{5}\un{G}$.

The friction-induced instability of toroidal flux tubes is also 
relevant in other astrophysical contexts, such as in circumstellar
accretion discs.
Accretion of mass implies the transport of angular momentum to larger
radii, which requires an efficient coupling of adjacent disc annuli.
The coupling of Keplerian shear flow through weak magnetic fields gives
rise to the magneto-rotational instability (MRI), which generates
magnetic turbulence and efficiently increases the viscous coupling and
angular momentum transport \citep{1991ApJ...376..214B}.
The MRI requires a poloidal magnetic field and is independent of the
toroidal field component.
This axi-symmetric shear instability occurs for vertical wavelengths
larger than a critical, field-dependent value, though growth rates and
threshold are independent of field strength.
It is an interesting possibility that the friction-induced instability of 
toroidal fields complements the MRI by supporting the radial magnetic
coupling of a disc.
Both instability mechanisms produce their highest growth rates on large
length scales, yet the friction-induced instability does intrinsically
not saturate\footnote{Since the critical wave length, beyond which the
MRI occurs, increases with the field strength, it can exceed the
vertical extend of the disc if the field strength is sufficiently
high.} at high field strengths as the MRI.
However, it is unclear what mechanisms aside of Keplerian shear can
produce the required velocity difference between the plasma inside and
outside the flux tube.

\citet{1996A&A...308.1013S} investigated the dynamics and evolution of
small flux rings in an accretion disc, including the drag force caused
by shear flows.
If a flux ring is deformed by a strong shear flow, variations of the
tension force and density contrast occur along the flux tube, which
produce two loops rising vertically through the disc until they emerge
into the disc corona.
These investigations confirm the strong influence of the drag force on
the dynamics of flux tubes with small minor tube radii, but they do not
consider friction-induced instabilities.
A stability analysis including the drag force would be possible,
because there is a flow component perpendicular to the axis of the
equilibrium flux tube.
Since differential rotation has a significant influence on the
stability properties of magnetic flux rings
\citep[e.g.][]{1995GAFD...81..233}, it will be necessary to include the
effect of shear flows and more general equilibrium conditions in the
stability analysis of the friction-induced instability in the framework
of accretion discs.

It is not clear to which extend the results based on a linear
velocity dependence of friction carry over to the non-linear case.
Given the $\alpha$-independence of the instability thresholds, the actual 
functional form of friction, that is, linear or quadratic, seems to be
of secondary importance, provided that it is anti-parallel to the
flow velocity perpendicular to the tube axis.
Numerical simulations based on the drag force confirm the properties of
friction-induced instability derived here.
A detailed analysis of the non-linear case and parameter study is the
topic of a forthcoming paper in this series.

\section{Conclusion}
\label{conc}
The friction-induced instability determines the stability properties of 
flux rings at the bottom of the convection zone, as its threshold is at 
lower field strengths than those of buoyancy- and tension-driven
instabilities.
The stability properties depend on the magnetic field strength
\emph{and} on the magnetic flux.
The threshold also depends on the location of the flux ring,
since the toroidal geometry causes a connection between the instability 
criterion and the equilibrium condition.
The analytical results confirm that a magnetic flux tube subject to
friction becomes unstable when a backward wave occurs, that is, when an
eigenmode reverses its direction of propagation.
We expect that friction-induced instabilities are also relevant in other
astrophysical contexts, such as the angular momentum transport in
accretion discs.

\begin{acknowledgements}
The author thanks Manfred Sch\"ussler, Robert Cameron, and Dieter
Schmitt for fruitful discussions and helpful comments.
\end{acknowledgements}

\bibliographystyle{aa}
\bibliography{}

\begin{appendix}

\section{Linear perturbations}
\label{pert}
We summarise the expressions describing adiabatic perturbations of
toroidal magnetic flux tubes in mechanical equilibrium.
Their detailed derivation can be found, for example, in
\citet{1998GApFD..89...75S} and \citet{1995GAFD...81..233}.

Small displacements, $\vec{\xi} (s,t)= \left( \xi_t, \xi_n, \xi_b
\right)$, of the flux tube from its equilibrium position change the
tangential, normal, and binormal vectors, $\vec{t},\vec{n},\vec{b}$,
respectively, of the Frenet basis, the curvature, $\kappa$, the flow
velocity, $u$, and the acceleration of the internal plasma:
\beqa
\vec{t}_1 
&=& 
\left( 
  \xi_{n,s} 
  + 
  \kappa_0 
  \xi_{t} 
\right) 
\vec{n}_0  
+ 
\xi_{b,s} 
\vec{b}_0  
\label{pert:t1}
\\
\vec{n}_1 
&=& 
- 
\left( 
  \xi_{n,s} 
  + 
  \kappa_0 
  \xi_{t} 
\right) 
\vec{t}_0 
+ 
\kappa_0^{-1}
\xi_{b,ss} 
\vec{b}_0  
\label{pert:n1}
\\
\vec{b}_1 
&=& 
- 
\xi_{b,s} 
\vec{t}_0  
- 
\kappa_0^{-1}
\xi_{b,ss} 
\vec{n}_0 
\label{pert:b1}
\\
\kappa_1
& =  &
\xi_{n,ss} 
+ 
\kappa_0^2
\xi_{n}
\label{pert:kap1}
\\
\vec{u}_1 
& = & 
\vec{t}_0 
\left( 
  \xi_{t,t} 
  + 
  u_0 
  \xi_{t,s} 
  - 
  u_0 
  \kappa_0 
  \xi_{n} 
\right) 
\nonumber \\ & & {}
+
\vec{n}_0 
\left( 
  \xi_{n,t} 
  + 
  u_0 
  \xi_{n,s} 
  + 
  u_0 
  \kappa_0 
  \xi_{t} 
\right) 
\nonumber \\ & & {}
+
\vec{b}_0 
\left( 
  \xi_{b,t} 
  + 
  u_0 
  \xi_{b,s} 
\right)
\label{pert:v1}
\\
\left( 
 \frac{d}{d t} \vec{u} 
\right)_1 
& = & 
\vec{t}_0 
\left(
  \xi_{t,tt} 
  + 
  2 
  u_0 
  \xi_{t,st} 
  - 
  2 
  u_0 
  \kappa_0 
  \xi_{n,t} 
  + 
  u_0^2 
  \xi_{t,ss} 
\right. \nonumber \\ & & {} \left. \qquad
  - 
  2 
  u_0^2 
  \kappa_0 
  \xi_{n,s}
  - 
  u_0^2 
  \kappa_0^2 
  \xi_{t}
\right)
\nonumber \\ & & {}
+ 
\vec{n}_0 
\left(
  \xi_{n,tt} 
  + 
  2 
  u_0 
  \xi_{n,st} 
  + 
  2 
  u_0 
  \kappa_0 
  \xi_{t,t} 
  + 
  2 
  u_0^2 
  \kappa_0 
  \xi_{t,s} 
\right. \nonumber \\ & & {} \left. \qquad
  - 
  u_0^2 
  \kappa_0^2 
  \xi_{n} 
  + 
  u_0^2 
  \xi_{n,ss} 
\right)
\nonumber \\ & & {}
+ 
\vec{b}_0 
\left(
  \xi_{b,tt} 
  + 
  2 
  u_0 
  \xi_{b,st} 
  + 
  u_0^2 
  \xi_{b,ss}
\right)
\label{pert:a1}
\ .
\eeqa
Quantities with index '0' and '1' are equilibrium and perturbed values,
respectively, and indices `$,t$' and `$,s$' describe partial
derivatives with respect to time, $t$, and equilibrium arclength,
$s_0$.

The flux tube is in pressure equilibrium with its environment.
Assuming adiabatic perturbations, the field strength, density, and
density contrast, $\Delta \rho= \rho - \rho\dw{e}$, change according to
the following equation:
\beqa
\frac{B_1}{B_0} 
& = &
\frac{\gamma \beta}{2+\gamma\beta} 
\frac{\left( \vec{g} \cdot \vec{\xi} \right)}{c\dw{s,0}^2}
+ 
\frac{\gamma\beta}{2+\gamma\beta}
\left( \xi_{t,s} - \kappa_0 \xi_{n} \right)
\label{pert:b10}
\\
\frac{\rho_1}{\rho_0}
& = &
\frac{\gamma \beta}{2+\gamma\beta} 
\frac{\left( \vec{g} \cdot \vec{\xi} \right)}{c\dw{s,0}^2}
- 
\frac{2}{2+\gamma\beta}
\left( \xi_{t,s} - \kappa_0 \xi_{n} \right)
\label{pert:rho1}
\\
\frac{\Delta\rho}{\rho}\bigg|_1
& = &
\frac{2}{\gamma \beta}
\Delta
\frac{\left( \vec{g} \cdot \vec{\xi} \right)}{c\dw{s,0}^2}
- 
\frac{2}{2+\gamma\beta}
\left( \xi_{t,s} - \kappa_0 \xi_{n} \right)
\label{pert:delrho1}
\ ,
\eeqa
with $\beta= 8\pi p_0 / B_0^2$ and 
\beq
\Delta= \frac{\beta}{1+\beta}
\left[
 \beta \delta 
 - 
 \frac{2}{\gamma} \left( \frac{1}{\gamma} - \frac{1}{2} \right) 
 \frac{\gamma\beta}{2+\gamma\beta} 
\right]
\ .
\eeq
The superadiabaticity $\delta= \nabla - \nabla\dw{ad}$ quantifies the
stability of the stratification against convective turnover ($\delta>
0$: convectively unstable).
From the integrated Walen equation, $\frac{B}{\rho}= \frac{B_0}{\rho_0}
\frac{\partial s}{\partial s_0}$, follows
\beq
\left(
  \frac{1}{4\pi} 
  \frac{B}{\rho} 
  \frac{\partial}{\partial s} B 
\right)_1 
= 
\frac{B_0^2}{4\pi\rho_0} 
\left( \frac{B_1}{B_0} \right)_{,s}
\ .
\label{pert:walen}
\eeq

\section{Coefficients of dispersion polynomials}
\label{coefs}

\subsection{General case}
\label{coeffsgeneral}
Based on the linearised equations of motion in the form of Eq.\
(\ref{algsys}), the eigenfrequencies of magnetic flux rings in
mechanical equilibrium parallel to the equatorial plane are determined
by the monic dispersion polynomial
\beq
\sum\limits_{j=0}^6
c_j
\tilde{\omega}^j
=
0
\ ,
\label{dispgen}
\eeq
which stems from the coefficient matrix of Eq.~(\ref{algsys}).
The coefficients are:
\beqa
c_5
& = &
-
6 m Ro
+
2 i \tilde{\alpha}
\label{dc5gen}
\\
c_4
& = &
4 \left(3 m^2-1\right) Ro^2
-
2 \left(3 m^2+4\right) Ro
-
4
\nonumber \\ & & {}
-
g_\perp^2 \tilde{\omega}\dw{MBV}^2
+
2 \frac{g\dw{n}}{\gamma f_0} \eta^2
-
8 i \tilde{\alpha} m Ro
-
\tilde{\alpha}^2
\label{dc4gen}
\\
c_3
& = &
-
8 m \left(m^2-1\right) Ro^3
+
8 m \left(3 m^2+1\right) Ro^2
\nonumber \\ & & {}
+
4 m g_\perp^2 \tilde{\omega}\dw{MBV}^2 Ro
-
4 m \frac{g\dw{n}}{\gamma f_0} \left(Ro-1\right) \eta^2
\nonumber \\ & & {}
+
i \tilde{\alpha} 
\big[
 4 \left(2 m^2-1\right) Ro^2
 -
 8 \left(m^2+1\right) Ro
 -
 4
\nonumber \\ & & \quad {}
 -
 g_\perp^2 \tilde{\omega}\dw{MBV}^2
 +
 2 \frac{g\dw{n}}{\gamma f_0} \eta^2
\big]
+
2 \tilde{\alpha}^2 m Ro
\label{dc3gen}
\\
c_2
& = &
-
24 m^2 \left(m^2-1\right) Ro^3
+
4 m^2 \left(3 m^2+7\right) Ro^2
+
8 m^2 Ro 
\nonumber \\ & & {}
-
4 
\tilde{\omega}\dw{MBV}^2 
\big[
 g_\perp^2 m^2 \left(Ro-1\right) Ro
 -
 g\dw{b}^2 \left(Ro+1\right)^2
\big]
\nonumber \\ & & {}
-
12 m^2 \frac{g\dw{n}}{\gamma f_0} Ro \eta^2
-
\left( \frac{1}{\gamma f_0} \right)^2 
\left(g_\perp^2 m^2+g\dw{b}^2\right) \eta^4
\nonumber \\ & & {}
+
2 i \tilde{\alpha} m 
\big[
 4 \left(2 m^2-1\right) Ro^2
 -
 4 Ro
\nonumber \\ & & \quad {}
 +
 g_\perp^2 \tilde{\omega}\dw{MBV}^2 Ro
 +
 2 \frac{g\dw{n}}{\gamma f_0}
 \eta^2
\big]
+
2 \tilde{\alpha}^2 m^2 Ro
\label{c2gen}
\\
c_1
& = &
-
24 m^3 \left(m^2-1\right) Ro^3
+
16 m^3 Ro^2
\nonumber \\ & & {}
-
8 m \tilde{\omega}\dw{MBV}^2 
\left[
 \left(g_\perp^2 m^2-g\dw{b}^2\right) Ro
 -
 g\dw{b}^2
\right]
 Ro
\nonumber \\ & & {}
-
8 m^3 \frac{g\dw{n}}{\gamma f_0} Ro \eta^2
+
2 m \left(\frac{1}{\gamma f_0}\right)^2
\left[
 \left(g_\perp^2 m^2-g\dw{b}^2\right) Ro
 -
 2 g\dw{b}^2
\right]
\eta^4
\nonumber \\ & & {}
+
i \tilde{\alpha} m^2
\big[
 4\left(2 m^2-1\right) Ro^2
 +
 2 g_\perp^2 \tilde{\omega}\dw{MBV}^2 Ro
 -
 \left(\frac{g_\perp}{\gamma f_0}\right)^2 \eta^4
\big]
\label{c1gen}
\\
c_0
& = &
-
8 m^4 \left(m^2-1\right) Ro^3
\nonumber \\ & & {}
-
2 m^2 
\left(g_\perp^2 m^2-g\dw{b}^2\right) 
\left[
 2 \tilde{\omega}\dw{MBV}^2 Ro
 -
 \left( \frac{1}{\gamma f_0} \right)^2 \eta^4
\right] 
Ro
\eeqa

\subsection{Axi-symmetric perturbations}
\label{coeffsaxisym}
In the special case of axial symmetric perturbations the dispersion
equation reduces to the monic quartic polynomial in Eq.\
(\ref{axipertdispeq}).
The friction-independent part of the coefficients are:
\beqa
c_2
& = &
2
\left( \frac{g\dw{n}}{\gamma f_0} - 2 \right) Ro \left( Ro + 2 \right)
-
4
-
g_\perp^2\,\tilde{\omega}\dw{MBV}^2
\label{c2tori}
\\
c_0
& = &
g\dw{b}^2 
\left(
4 \tilde{\omega}\dw{MBV}^2 \left(Ro+1\right)^2
-
\left( \frac{1}{\gamma f_0} \right)^2
Ro^2 \left(Ro+2\right)^2
\right)
\label{c0tori}
\eeqa

\subsection{Flux tubes in the equatorial plane}
\label{coeffsequa}
For flux rings in the equatorial plane the dispersion equation
factorises in a quadratic and quartic polynomial, which describe the
perturbation perpendicular and parallel to the equatorial plane,
respectively.
The coefficients of the latter are:
\beqa
c_3
& = &
-
4\,m\,Ro
+
i\,\tilde{\alpha}
\label{c3equa}
\\
c_2
& = &
4\,\left(m^2-1\right)\,Ro^2
-
4\,\left(m^2+2\right)\,Ro
-
4
\nonumber \\ & & 
{}
-
\tilde{\omega}\dw{MBV}^2\,g_n^2
+
2\,\frac{g_n}{\gamma f_0}\,\eta^2
-
2\,i\,\tilde{\alpha}\,m\,Ro
\label{c2equa}
\\
c_{1}
& = &
8\,\left(m^2-1\right)\,m\,Ro^2
-
8\,m\,Ro
\nonumber \\ & & 
{}
+
2\,m\,g\dw{n}^2\,\tilde{\omega}\dw{MBV}^2\,Ro
+ 
4\,m\,\frac{g\dw{n}}{\gamma f_0}\,\eta^2
-
2\,i\,\tilde{\alpha}\,m^2\,Ro
\label{c1equa}
\\
c_0
& = &
4\,\left(m^2-1\right)\,m^2\,Ro^2
\nonumber \\ & & {}
+
2\,m^2\,g\dw{n}^2\,\tilde{\omega}\dw{MBV}^2\,Ro
-
m^2\,\left( \frac{g\dw{n}}{\gamma f_0} \right)^2\,\eta^4
\label{c0equa}
\eeqa

\end{appendix}

\end{document}